\documentclass[apjl]{emulateapj}
\usepackage{revsymb}
\usepackage{graphicx}
\usepackage{amsmath,amssymb}
\usepackage{xcolor}

\begin{document}

\title{VVV Survey Observations of a Microlensing Stellar Mass Black Hole Candidate in the Field of the Globular Cluster NGC\,6553}

\author{D. Minniti$^{1,2,3}$,
R. Contreras Ramos$^{2,4}$,
J. Alonso-Garc\'ia$^{2,8}$,
T. Anguita$^{1,2}$,
M. Catelan$^{2,4}$,
F. Gran$^{4,2}$,
V. Motta$^{5}$,
G. Muro$^{6}$,
K. Rojas$^{5}$, \&
R. K. Saito$^{7}$}

\affiliation{$^{1}$Departamento de Ciencias Fisicas, Universidad Andres Bello, Campus La Casona, Fern\'andez Concha 700, Santiago, Chile}; \email{email:dante@astrofisica.cl}
\affiliation{$^{2}$Millennium Institute of Astrophysics, Av. Vicu\~na Mackenna 4860, 782-0436 Macul, Santiago, Chile }
\affiliation{$^{3}$Vatican Observatory, Vatican City State, I-00120, Italy}
\affiliation{$^{4}$Instituto de Astrof\'isica, Pontificia Universidad Cat\'olica de Chile, Av. Vicu\~na Mackenna 4860, 782-0436 Macul, Chile }
\affiliation{$^{5}$Instituto de F\'isica y Astronom\'ia, Facultad de Ciencias, Universidad de Valpara\'iso, Av. Gran Breta\~na 1111, Playa Ancha,  Valpara\'iso, Chile }
\affiliation{$^{6}$Instituut voor Sterrenkunde, K.U.Leuven, Celestijnenlaan 200B, B-1 Heverlee, Belgium}
\affiliation{$^{7}$Universidade Federal de Sergipe, Departamento de F\'isica, Av. Marechal Rondon s/n, 49100-000, S\~ao Crist\'ov\~ao, SE, Brazil}
\affiliation{$^{8}$ Unidad de Astronom\'ia, Facultad Cs. B\'asicas, Universidad de Antofagasta, Avda. U. de Antofagasta 02800, Antofagasta, Chile}

\begin{abstract}
We report the discovery of a large timescale candidate microlensing event of a bulge stellar source based on near-infrared observations with the VISTA Variables in the V\'ia L\'actea Survey (VVV). The new microlensing event is projected only 3.5 arcmin away from the center of the globular cluster NGC\,6553. The source appears to be a bulge giant star with magnitude $K_{\rm s}=13.52$, based on the position in the color-magnitude diagram. The foreground lens may be located in the globular cluster, which has well-known parameters such as distance and proper motions. If the lens is a cluster member, we can directly estimate its mass simply following \cite{paczynski96} which is a modified version of the more general case due to Refsdal. In that case, the lens would be a massive stellar remnant, with $M=1.5-3.5 M_\odot$. If the blending fraction of the microlensing event appears to be small, and this lens would represent a good isolated black hole (BH) candidate, that would be the oldest BH known. Alternative explanations (with a larger blending fraction) also point to a massive stellar remnant if the lens is located in the Galactic disk and does not belong to the globular cluster.
\end{abstract}

\keywords{globular clusters: individual (NGC\,6553) -- gravitational lensing: micro -- infrared: stars -- surveys}

\section{Introduction}
Microlensing is the best (and so far the only) tool to detect isolated stellar mass black holes \citep[BHs; see][]{mao12}. The microlensing projects have discovered a few candidate stellar-mass BH lenses toward the Galactic bulge \citep{bennett02,mao12,agol02,poindexter05}. \cite{gould99} specifically applied astrometric microlensing in order to measure the masses of BHs. Although the measurements remain challenging, this method is important for future measurements from space \citep{gould14}.

\cite{refsdal64} showed that the mass $M$ of a gravitational lens is related to observables ($\mu_{rel}, \pi_{rel}, t_{E}$) by:
\begin{equation}
M=\frac{\theta^2_{E}} {\kappa \pi_{rel}}, ~\theta_{E} = \mu_{rel}t_{E}, ~\kappa \equiv \frac{4G}{c^2 AU} \simeq 8.1 \frac{mas}{M_{\odot}}
\end{equation}
where $t_{E}$ is the Einstein crossing time derived from the event light curve and $\pi_{rel}$ and $\mu_{rel}$ are the lens-source relative parallax and proper motion respectively. While \cite{refsdal64} clearly had in mind that the lens and source could both be seen, \cite{paczynski94} pointed out that for globular-cluster lenses detected via their magnification of Galactic-bar sources, $\pi_{rel}$ and $\mu_{rel}$ can potentially be estimated without directly observing the lens. This opens the possibility that the method could be applied to BH lenses.
The main advantage of such a search is that the microlensing geometry is more certain, and there is a direct relationship between the lens mass and the time scale of the event. This technique was successful in the case of the first confirmed globular cluster microlensing event in M22 (NGC\,6656) due to a low-mass star \citep{pietrukowicz05,pietrukowicz12}. Later, another microlensing event in close projection to the globular cluster NGC\,6553 was reported \citep{yee13}, which also turned out to be a low-mass lens, possibly with a planet. However, the suggestion by \cite{paczynski94,paczynski96} seems to be a promising way to search and measure heavy remnants (neutron stars (NSs), or BHs) in spite of the technical challenges.

Here we report the discovery of a long-timescale candidate microlensing event of a bulge stellar source projected only 3.5 arcmin away from the center of the globular cluster NGC\,6553. If lens membership to the cluster is confirmed, the lens could be a massive stellar remnant, possibly a BH with $\sim M=2M_\odot$. Section 2 presents the VVV Survey observations of the microlensing event. Section 3 discusses the globular cluster NGC\,6553 where the lens may reside. Section 4 shows the measured parameters of the lens, and Section 5 summarises the conclusions and future tests.

\section{VVV Survey Observations of the Microlensing Event}
VISTA Variables in the V\'ia L\'actea (VVV) is a public ESO near-infrared (IR) variability survey aimed at scanning the inner Milky Way \citep{minniti10}. The observations are acquired with the VISTA 4 m telescope at ESO Paranal Observatory \citep{sutherland15}. The VVV survey covers an area of 562 sqdeg in the inner disk and bulge of the Milky Way. The VVV database now contains multicolor ($ZYJHK_{\rm s}$) photometry, and multiple epochs in the $K_{\rm s}$-band, monitoring about a billion sources in total \citep{saito12}. The $K_{\rm s}$-band observations continue, and the variability light curves so far span from 2010 to 2015. This large database enables a number of studies of different variable objects, including microlensing events \citep{catelan13,hempel14}.

We have started to search for variable stars using the VVV Survey data in the fields of 36 globular clusters toward the Galactic bulge. The data reductions and photometry have been described in detail by \cite{alonsogarcia15}, where our first results for the globular clusters 2MASS-GC02 and Terzan 10 are also presented. Briefly, the individual VVV Survey images are reduced, astrometrized and stacked by the Cambridge Astronomy Survey Unit (CASU) using the VISTA Data Flow System (VDFS) pipeline \citep{emerson04,irwin04,hambly04}, and the photometry is calibrated onto the VISTA filter system. Then we carry out point-spread function (PSF) photometry on the individual processed stacked images using an updated version of DoPHOT \citep{schechter93,alonsogarcia12}. The photometry for each object in the different images available was cross-correlated using the STILTS package \cite{taylor06}, and the light curves generated were then analyzed for variability \citep[see][]{alonsogarcia15}.

During this near-IR search for variable stars in the field of the globular cluster NGC\,6553 \citep[][in preparation]{contrerasramos15}, we discovered a candidate microlensing event located at R.A.(2000)$=$18:09:13.86, decl.(2000)$=-$25:57:52.7, that is only 3.5 arcmin away from the cluster center. This event peaked in the 2012 bulge season (on the 2012, July 8), and it has not been reported by any of the ongoing microlensing surveys. The brightness of the object increased by about 0.7 mag in the $K_{\rm s}$-band over 50 days, reaching a magnitude of $K_{\rm s}=12.8$, and then fading to a constant level of $K_{\rm s}=13.5$. Table 1 shows the $K_{\rm s}$-band observations for this object. The $K_{\rm s}$-band epochs correspond to different nights, and the seeing for the majority of them was very good, between 0.6 and 1.2 arcsec.

\begin{table}
  \centering
  \caption{VVV Survey near-IR photometry of the candidate microlensing event.}
  \begin{tabular}{ccc}
    \hline
    Epoch (MJD) & $K_{\rm s}$ & $\sigma_{\rm{K}_{\rm s}}$ \\
    \hline
    56363.376890 & 13.509 & 0.013 \\
    55298.333635 & 13.509 & 0.015 \\
    55491.015648 & 13.528 & 0.013 \\
    55778.139393 & 13.495 & 0.013 \\
    55804.103650 & 13.543 & 0.012 \\
    \hline \hline
  \end{tabular}
  \tablecomments{Table 1 is published in its entirety in the electronic edition of
   the Astrophysical Journal. A portion is shown here for guidance regarding its
  form and content.}
\end{table}

Fig.~1 shows the $6.8\times3.5$ arcmin finding chart of the microlensing event with respect to NGC\,6553. This lies well within the cluster tidal radius of $R=8.16$ arcmin. The zoomed $30\times30$ $\rm{arcsec}^2$ region to the right shows that the source is located just in between two brighter stars (both saturated with $K_{\rm s}<12$) that are separated by 3 arcsec. This makes the photometry difficult in the worst-seeing images, explaining a few of the deviant points with large error bars in the light curve (Fig.~2). 

\begin{figure*}
\centering
\includegraphics[scale=0.6]{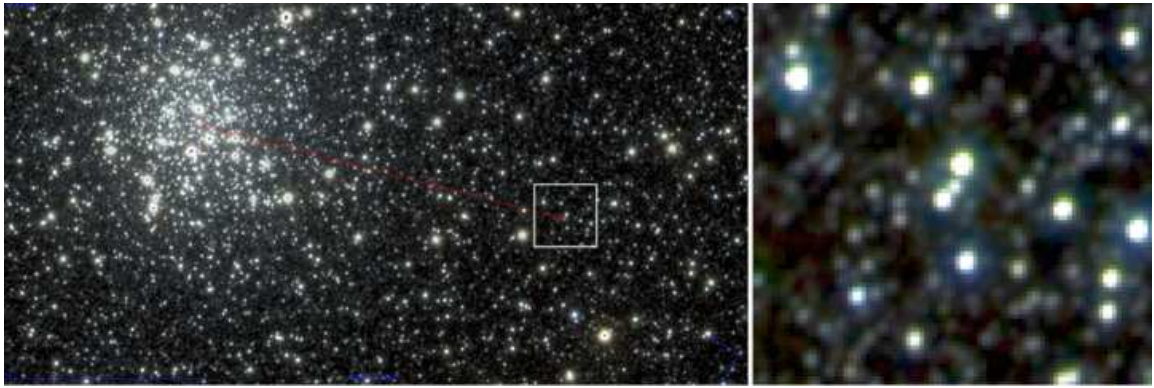}
\caption{ Left: $6.8\times3.5$ arcmin finding chart showing the position of the microlensing event with respect to NGC\,6553. The projected distance to the cluster center is 3.5 arcmin, well within the cluster tidal radius of $R=8.16$ arcmin. Right: zoomed $30\times30$ $\rm{arcsec}^2$ region centered on the source star, showing that it is located just in between two brighter stars (with $K_{\rm s}<12$) that are separated by 3 arcsec. The seeing of the images is $0.8^{\prime\prime}$}
\end{figure*}

The source appears unblended in the best seeing images (we have also checked in the {\it Hubble Space Telescope (HST)} archive and unfortunately there is no image of this field), and Fig.~2 shows the light curve for this event, along with the simple unblended microlensing fit \citep{paczynski94,paczynski96}. This would correspond to an event without significant contamination from an unresolved source. However, we have fitted a microlensing light curve, obtaining two equally good fits with $\chi^2=200$: an unblended and a blended fit, that are discussed in turn in the following sections. Both fits are very good, with the following microlensing parameters: Case A (without blending): baseline magnitude $K_{\rm{s,0}}=13.515\pm0.002$, impact parameter $u_0= 0.62\pm0.004$, time of closest approach $t_{0}=56117.5\pm0.43$, Einstein timescale $t_{E}=51.3\pm0.8$ days, blending $f=1.00$, and $\chi^2=201$. Case B (with blending): baseline magnitude $K_{\rm{s,0}}=13.515\pm0.002$, impact parameter $u_{0}=0.46\pm0.08$, time of closest approach $t_{0}=56117.4\pm0.42$, Einstein timescale $t_{E}=62.5\pm9$ days, blending $f=0.61\pm0.17$, and $\chi^2=199$.

\begin{figure}
\centering
\includegraphics[scale=0.45]{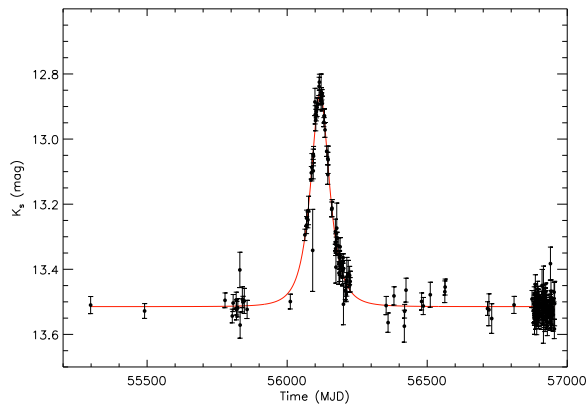}
\caption{$K_{\rm s}$-band light curve for this event, along with a simple microlensing fit \citep{paczynski94}. The microlensing parameters of this fit are: baseline magnitude $K_{\rm s,0}=13.515\pm0.002$, impact $u_0=0.62\pm0.004$, time of closest approach $t_{0}=56117.5\pm0.43$, Einstein timescale $t_{E}=51.3\pm0.8$, blending fraction $f=1.00$, and $\chi^2=200$. The points with the larger errorbars correspond to the worst seeing images.}
\end{figure}

The values obtained for $\chi^2$ show that there is no improvement in the curve fitting when a blending factor is included. Also, the VVV Survey is designed to monitor variable stars, and the light curve is not frequently and evenly sampled, preventing us from using more complex models (with larger numbers of free parameters like those including the parallax effect) with better confidence than a simple microlensing light curve. 

Can this be a variable star instead of a microlensing event?~The main kinds of variable star contaminants are foreground dwarf novae and distant background SN, both of which are unlikely. The shape of the light curve is very symmetric, unlike the nova, dwarf novae or supernova light curves that typically show a fast rise and a slow decline. There are some exceptions, as some recurrent novae show symmetric outbursts (see for example, the light curve of GK Per\footnote{http://www.aavso.org/sites/default/files/images/LTGKper.gif}). However, we discard a dwarf nova outburst because of the light curve shape and also because our object is too red ($J-K_{\rm s}=1.0$) to be a dwarf novae. We also searched for variability in the region outside the peak but no signal of eclipses or ellipsoidal variations from a binary system were found as expected from a cataclysmic variable. Also, even though the VVV Survey data is deep enough to see background galaxies throughout the bulge, in this case no extended galaxy-like object that could be a supernova host galaxy is detected.

\section{The globular cluster NGC\,6553}

Because of its proximity to a globular cluster, this microlensing event is particularly interesting. The bulge globular cluster NGC\,6553 has been well studied, and its physical parameters are relatively well known.  NGC\,6553 is an old and
metal-rich bulge globular cluster \citep{ortolani95,minniti95,barbuy98,zoccali01,alvesbrito06}. This cluster is moderately concentrated with a core radius $r_c=0.55$ arcmin, and a tidal radius $r_t=8.16$ arcmin. It is very reddened, with E($B-V)=0.73$, A$_{\rm V}=2.26$, and A$_{\rm{K}_{\rm{s}}}=0.23$ \citep{barbuy98}. Particularly relevant for this work is the distance of the cluster. We adopt a distance of 6.0 kpc for this cluster following \cite{alvesbrito06}, as listed in the 2010 version of \cite{harris96} catalog, noting that the distance uncertainty adds to the total error budget. We observe the cluster horizontal branch (HB) at $K_{\rm s}=12.5$ (Fig.~3), which is consistent with this distance measurement.

\cite{zoccali01} measured a relative proper motion of NGC\,6553 with respect to the bulge of $\mu_{\ell}=5.89$ and $\mu_b=0.42$ mas/yr. This gives a relative mean proper motion difference between bulge and cluster stars in the sky of 5.9 mas/yr. That would be the expected mean relative proper motion of the bulge source and the lens if the latter is a globular cluster member. Of course, the large bulge velocity dispersion can result in a larger or smaller actual relative proper motion.

\begin{figure}
\centering
\includegraphics[scale=0.44]{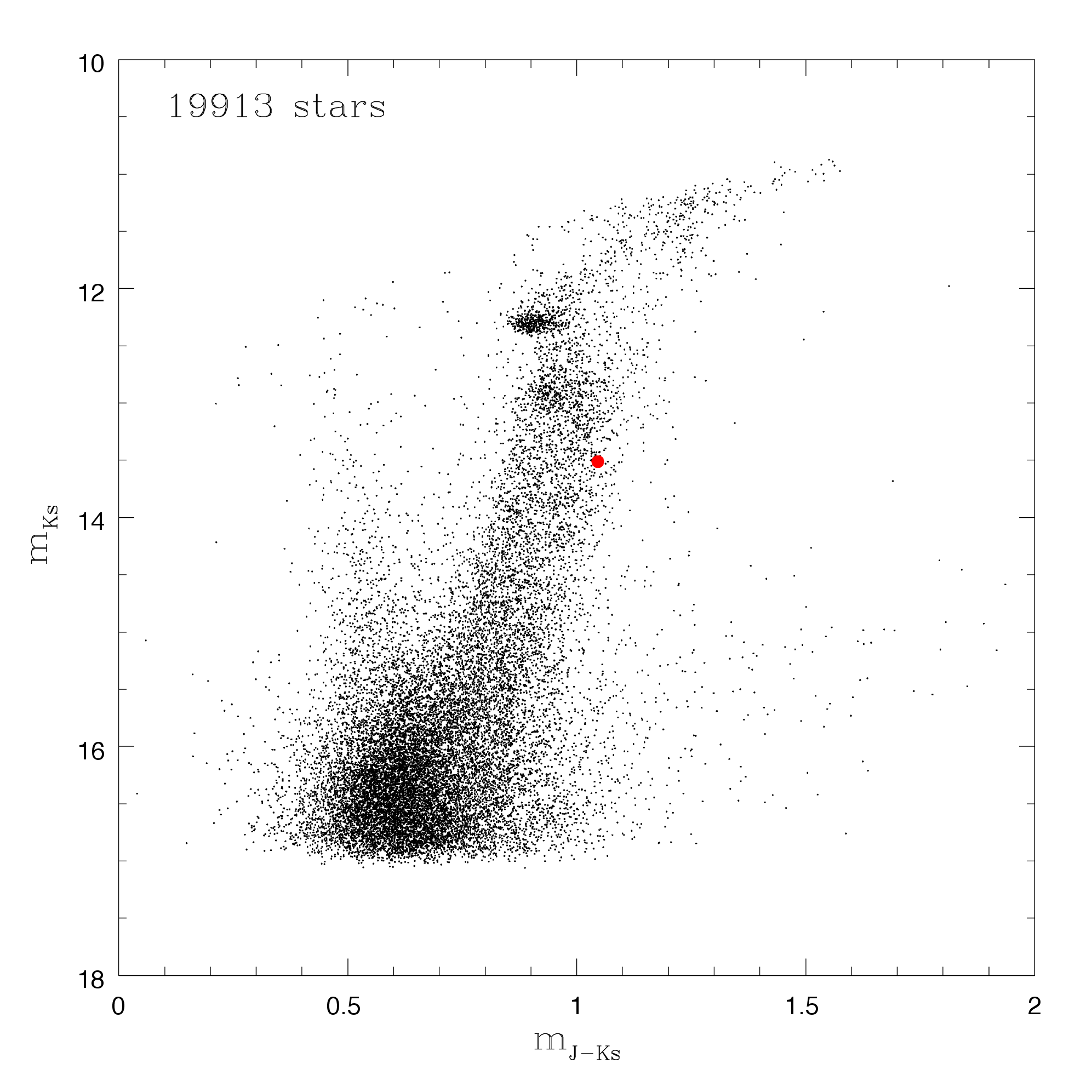}
\caption{$K_{\rm s}$ vs $J-K_{\rm s}$ near-infrared color-magnitude diagram (CMD) of about 20000 stars in $4\times4$ $\rm{arcmin}^2$ field centered on the globular cluster NGC\,6553. The red circle marks the source star of the microlensing event. This CMD shows the foreground disk main sequence, the populated globular cluster main-sequence (MS) turn-off, the cluster red giant branch (RGB), its red horizontal branch (HB), and its red giant bump, as well as a redder and wider background bulge RGB, including the bulge red clump. We caution that photometric non-linearity and saturation starts at $K_{\rm s}<12.5$.}
\end{figure}

The source star in particular (with $K_{\rm s}=13.5$, $J-K_{\rm s}=1.0$ at the baseline) is redder than the RGB of the globular cluster, consistent with a bulge giant. The source is also fainter than the bulk of the bulge red clump stars. If the source is a red clump giant star, it must be located on the far side of the bulge, at $D_s \sim 9$ kpc.

The field is very crowded and contains numerous bulge giants, and therefore membership of the lens to the cluster cannot be unambiguously secured. Luckily the VVV Survey images cover a large area (1.5 sqdeg), and we can measure the background contamination. At the distance of 3.5 arcmin from the center of the cluster, we estimate that roughly 20\% of the stars per unit area are cluster members. However, that may or may not apply to the distribution of heavy remnants. In contrast, the microlensing event discussed by \cite{yee13} was located 6.7 arcmin away from this cluster, where the cluster stellar density is 6\%, based on which they estimated that there is only 18\% probability that their lens is a cluster member. Following their procedure we estimate that there is roughly 50\% probability that the lens, located 3.5 arcmin from NGC\,6553, actually belongs to the cluster.

\section{The Candidate BH Lens}
The dark microlensing events toward the bulge (i.e. with no light from the lens) that have long-duration ($>100$~days) may be due to massive stellar remnants. However, the masses based on timescales are statistical mass estimates. Long-duration events can also be due to slow-moving, low-mass objects. This makes the present event particularly interesting, since an Einstein diameter crossing time of $\hat{t}=103$ days (\cite{bennett02}), may be due to a relatively massive lens, such as a heavy non-luminous remnant (NS or BH).

Because of the position of the source star in the CMD (Fig.~3) we can assume that it is located in the Galactic bulge. Under this assumption, we consider two scenarios: In the first scenario the lens is a member of the globular cluster NGC\,6553. In this case we can estimate the lens mass \citep[c.f.][]{paczynski94,paczynski96}, because we know the distance and the proper motion of the cluster.


We have fitted different possibilities considering different distances to the source star. The mass determination is uncertain due to the unconstrained $\mu_{rel}$, due also to the considerable bulge line of sight depth \citep{nataf13}. The most likely source distance is where the density of bulge stars peaks along the line of sight, i.e. $D_s=8$ kpc. In this case, assuming $D_L=6$ for the lens, we obtain a lens mass $M_L=3.5 \pm 0.1 M_\odot$ (using relative transverse velocity 220 km/s), or $M_L=2.0 \pm 0.1 M_\odot$ (using relative transverse velocity 168 km/s). If the source star is located in the far side of the bulge, at $D_s=9$ kpc, we obtain a smaller lens mass, $M_L=2.7 \pm 0.1 M_\odot$ and  $M_L=1.5 \pm 0.1 M_\odot$ for both cases, but still within the realm of the heavy remnant hypothesis. As discussed in Section 2, there are two equally good fits, one with a dark lens (case A with blending parameter $f=1.0$), and another one with a significant contribution from an unresolved source (case B with $f=0.62\pm0.20$). In the case of the first fit, if the lens is a member of NGC\,6553, we estimate $M=3.1 \pm 0.4 M_\odot$ and $M=1.8\pm0.2 M_\odot$ for both cases, where the uncertainty is driven by the source distance and transverse motions. This fit then rules out other stellar remnants such as white dwarfs, and is more consistent with a NS or BH stellar remnant.

Even though we claim priority for recognising this as a BH candidate, this event was independently found by OGLE \footnote{reported online at:\\ http://ogle.astrouw.edu.pl/ogle4/ews/2012/ews.html} as event OGLE-2012-BLG-0548. We would like to note that the reported OGLE microlensing parameters agree quite well with the ones derived solely from VVV photometry. 

\cite{bennett02} estimate that the mean mass for six microlensing parallax events that they study is 2.7 $M_\odot$, arguing that they are BH candidates because they surpass the Chandrasekhar mass of 1.4 $M_\sun$, being even larger than the allowed maximum NS mass of 2.0  $M_\odot$. Interestingly, the mass we measure for the lens studied here is similar to the mean mass estimated for the six long timescale microlenses studied by \cite{bennett02}.

This is an interesting case to study, because if the lens is so massive and old, it should be located close to the center of the cluster. It is unexpected to find a heavy stellar remnant at a distance of about 6 core radii from the cluster center so long after its formation. For example, the populations of millisecond pulsars and LMXBs in globular clusters tend to be among the most radially concentrated, and it is not clear why this BH did not sink into the globular cluster center through dynamical friction \cite[e.g.,][]{heinke05}.

As another point, we note that XMM source \#27 in NGC\,6553 from \cite{guillot11} located at R.A.(2000)$=$18:09:21.67, decl.(2000)$=-$25:57:31.6 is the closest X-ray source to the position of the microlensing event located about 2 arcmin away. Based on the existing data, however, it appears unlikely that XMM source \#27 corresponds to radiation from this candidate BH, as expected if it is accreting gas from the interstellar medium or the wind of a companion star \cite[e.g.,][]{agol02}.

In the second scenario that we consider, the lens is not a member of the globular cluster NGC\,6553. In this case the distance to the lens is unknown and its mass is not well constrain at all. 
However we can make some assumptions to visualize possible solutions. If we consider the distance of the bulge for the source, a long timescale can be obtained for a low-mass star as lens only if this lens is located at a few kpc from the Sun. For a source at a distance of $D_S=8$ kpc, and assuming that the lens is located at $D_L=3$, 4, and 5 kpc with a typical disk velocity of  $V=220$ km/seg we estimate: $M_3=2.8 M_\odot$, $M_4=2.6 M\odot$, $M_5=2.8 M_\odot$, respectively. Instead, for a source at a distance of $D_S=9$ kpc, and a lens at $D_L=3$, 4 kpc, and 5 kpc we estimate: $M_3=2.6 M_\odot$, $M_4=2.3 M_\odot$, $M_5=2.3 M_\odot$, respectively. 

This case would favor more the second fit, with significant blending from an unresolved source. However, this is interesting because the inferred masses are large for a typical disk main sequence star. If the lens is a main sequence star, it should be bright enough for detection at these intermediate distances, and also bluer than the observed color $J-K_{\rm s}=1.0$, making this scenario less likely.  A lens closer to the Sun than 3 kpc is unlikely because of the same argument: it would be bright enough and should have been observed.

\section{Conclusions}
We have re-discovered a microlensing event in the field of the globular cluster NGC\,6553. Because the distance of this cluster is well known ($D=6.0$ kpc), and the CMD suggests that the source is likely a bulge red giant at $D_S=8-9$ kpc, we can solve for the mass of the lensing object if we assume this is a cluster member. 
Simply using eq. 15 of \cite{paczynski96} results in a lens mass of $M=1.5-3.5 M_\odot$ consistent with a heavy remnant. 

There is another equally good fit to the microlensing light curve, with significant blending from an unresolved source. However, we note that combining the optical and IR data from OGLE and VVV, the microlensing color can be more precisely measured, and it appears to be essentially similar to the baseline color, meaning that the blending is really small, or alternatively that the lens has exactly the same color as the source. Proper motion measurements are warranted in this case, and combining VVV and OGLE data would greatly increase the precision of the parallax measurement and will be of great importance to better constrain the mass of this object. The alternative scenario of a more nearby disk main sequence star lensing a distant bulge giant cannot be ruled out, but also yields a rather massive lens (2.3-2.8 $M_\odot$). Even though we cannot rule out this possibility, it appears to be less likely because in this case the lens would be detectable as a bright blue star. 

We can make specific predictions if the lensing object is a cluster member: the lens should eventually move away from the source because the relative proper motion of this globular cluster is relatively large, and then be detected if it is a stellar object, or remain invisible if it is an isolated BH. This method is the best way to confirm the microlensing nature of the event \cite{pietrukowicz12}. In the present case, the source-lens separation can reach the 60 mas range within 10 year \citep{zoccali01}, and the lensing star should be detected with high-resolution images (with HST or AO cameras) if it is not a BH in the cluster. In this sence, immediate high resolution $K$-band measurements using AO are desirable in order to provide a reference image for subsequent detection of the lens.

If membership to the cluster is confirmed with follow-up observations, this would not only be the only known stellar mass BH in a globular cluster, an important object to validate stellar evolution theory, but also the oldest BH known. This discovery opens up some interesting questions, like what is the mass distribution of globular cluster BHs produced by the death of ancient massive stars, why this BH did not sink into the globular cluster center through dynamical friction, how many of these low-mass BHs can be present in globular clusters, and how much these objects contribute to the total mass budget of dark remnants in the Milky Way and the Universe.

\bigskip

\acknowledgements
We gratefully acknowledge use of data from the ESO Public Survey programme ID 179.B-2002 taken with the VISTA telescope, and data products from the Cambridge Astronomical Survey Unit, and funding from the BASAL CATA Center for Astrophysics and Associated Technologies PFB-06, the Millennium Institute of Astrophysics MAS from the Ministry of Economy, Development, and Tourism through Programa Iniciativa Cient\'ifica Milenio grant awarded to the Millennium Institute of Astrophysics (MAS) ICM P07-021-F, Proyecto FONDECYT Postdoctoral 3130320, 3130552, and Proyectos FONDECYT Regular 1130196, 1120741, and 1141141, FONDECYT Iniciation 11130630, CONICYT-PCHA Magíster Nacional 2014/2015-22141509, Doctoral Scholarship FIB-UV 2014, CONICYT-Gemini 32120033, FIC-R Fund, project 30321072 and CNPq/Brazil Projects 310636/2013-2 and 481468/2013-7. We warmly thank the ESO Paranal Observatory staff for performing the observations and Mike Irwin, Eduardo Gonzalez-Solares, and Jim Lewis at CASU for pipeline data processing support.

\end{document}